\documentclass[a4paper,twocolumn,
english,aps,pre,floatfix,groupedaddress,showpacs,nofootinbib]{revtex4-1}
\usepackage[T1]{fontenc}
\usepackage[latin1]{inputenc}
\usepackage{amsmath}
\usepackage{babel}
\usepackage{graphics}
\usepackage{amssymb}
\usepackage{dcolumn}


\begin{document}
\title
{Driven flow with exclusion and transport in graphene-like structures} 
\author {R. B. \surname{Stinchcombe}}
\email{r.stinchcombe1@physics.ox.ac.uk}
\affiliation{Rudolf Peierls Centre for Theoretical Physics, University of
Oxford, 1 Keble Road, Oxford OX1 3NP, United Kingdom}
\author {S. L. A. \surname{de Queiroz}}
\email{sldq@if.ufrj.br}
\author {M. A. G. \surname{Cunha}}
\email{magc@if.ufrj.br}
\author {Belita \surname{Koiller}}
\email{bk@if.ufrj.br}
\affiliation{Instituto de F\'\i sica, Universidade Federal do
Rio de Janeiro, Caixa Postal 68528, 21941-972
Rio de Janeiro RJ, Brazil}

\date{\today}

\begin{abstract} 
We study driven flow with exclusion in graphene-like structures.
The totally asymmetric simple exclusion process (TASEP), a 
well-known model in its strictly one-dimensional (chain) version, is
generalized to cylinder (nanotube) and ribbon (nanoribbon) geometries.
A mean-field theoretical description is given for very narrow
ribbons ("necklaces"), and nanotubes. For specific configurations of 
bond transmissivity rates, and for a variety of boundary conditions, 
theory predicts equivalent steady state behavior between (sublattices on) 
these structures and chains. This is verified by numerical simulations,
to excellent accuracy, by evaluating steady-state currents. We
also numerically treat ribbons of general width. 
We examine the adequacy of this model to the description of 
electronic transport in carbon nanotubes and nanoribbons, or 
specifically-designed quantum dot arrays.
\end{abstract}
\maketitle
 
\section{Introduction} 
\label{intro} 

The impact of geometric and topological aspects of the atomic 
arrangements of materials on its electronic properties has been 
recognized for quite some time~\cite{Kittel}. A remarkable example is carbon 
(C), for which different bonding and 
valence states of C result in stable configurations of C-only based 
materials in all dimensions D, namely 3D - diamond, graphite, amorphous C, 
2D - graphene, 1D - nanotubes (CNT), nanoribbons (CNR), and 0D - 
fullerenes~\cite{RMP}. Except for diamond,
ordered C-based materials of all dimensionalities are constituted of 
stacked, deformed, or fragmented 2D graphene, which may thus be 
considered as the basic building block of all forms.

Here we use a simple model to  investigate transport properties on the
hexagonal geometries of the CNT and CNR structures. These systems are widely
accepted as being 1D, based on aspect ratio criteria. For processes such as
current flow, e.g. having bias and collective aspects analogous to those from
Coulomb blockade, one can question how the geometry affects the behavior
and in particular ask whether the physical quantity of interest in the
system is indeed equivalent to what is expected from a 
{\em bona fide} 1D system. 
The honeycomb structure of graphene implies a topology very 
different from a genuine 1D linear atomic array, where a one-to-one 
correspondence of bonds and atoms is trivially given; as seen in 
Sec.~\ref{sec:theory}, it requires introducing additional parameters in the
model used here.

We do not attempt a realistic description of electronic transport 
in C allotropes under an applied bias, which 
requires quantum mechanical description of the electrons in 
the respective ordered structure potential, as presented, e.g.,
in Refs.~\onlinecite{Caio_1,Caio}. 
Instead, we take a complementary viewpoint by generalizing a very simple transport 
model, extensively studied in 1D lattices, to graphene-like
nanotube and nanoribbon structures. 
This highlights the effect of the topology of the 
underlying skeleton on transport, allowing direct and unambiguous 
comparison between such systems and the well-established linear chain 
results obtained within the same model, namely the totally asymmetric 
simple exclusion process 
(TASEP)~\cite{derr98,sch00,mukamel,derr93,rbs01,be07,cmz11}. 
We find that within certain plausible 
assumptions nanotubes can be close to exact realizations of 1D systems while, 
surprisingly, narrow nanoribbons  deviate substantially from 1D 
behavior, which is however obtained only in the limit of very wide ribbons.

The TASEP is among the simplest models in non-equilibrium physics, while at the 
same time exhibiting 
many non-trivial properties including flow phase changes, because of its
collective character~\cite{derr98,sch00,mukamel,derr93,rbs01,be07,cmz11}. 
The TASEP and its generalizations have been
applied to a broad range of non-equilibrium physical contexts, from
the macroscopic level such as highway traffic~\cite{sz95} to the microscopic,
including sequence alignment in computational biology~\cite{rb02}
and current shot noise in quantum-dot chains~\cite{kvo10}.
 
In the time  evolution  of the $1+1$ dimensional TASEP,
the particle number $n_\ell$ at lattice site $\ell$ can be $0$ or $1$, 
and the forward hopping of particles is only to an empty adjacent site. 
In addition to the stochastic character provided by random selection of site occupation 
update~\cite{rsss98,dqrbs08}, the instantaneous current $J_{\ell,\ell+1}$ 
across the bond from $\ell$ to $\ell +1$ depends also on the stochastic attempt rate, 
$p_\ell$, associated with it. Thus,  
\begin{equation}
J_{\ell,\ell+1}= \begin{cases}{n_\ell (1-n_{\ell+1})\quad {\rm with\ probability}\ p_\ell}\cr
{0\qquad\qquad\qquad {\rm with\ probability}\ 1-p_\ell\ .}
\end{cases}
\label{eq:jinst}
\end{equation}
In Ref.~\onlinecite{kvo10} it was argued that the ingredients of TASEP
are expected to be physically present in the description of electronic
transport on a quantum-dot chain; namely,
the directional bias would be provided by an external voltage difference
imposed at the ends of the system, and the exclusion effect by on-site Coulomb 
blockade.

Here we exploit the consequences of applying a similar scenario to graphene-like
geometries. 
In Section~\ref{sec:theory} a general mean-field theoretic approach is developed, for 
the problem of driven flow with exclusion in two-dimensional structures which are
cutouts [$\,$"necklaces" (to be defined below), or cylinders, or ribbons etc$\,$] 
from a honeycomb lattice. Fundamental relationships, like that between the 
steady-state current $J$ and (i) the 
(site-averaged) particle density  (for periodic boundary conditions [$\,$PBC$\,$]) 
or (ii) the injection/ejection
parameters $\alpha$, $\beta$ (for systems with open ends) are 
given. Density profiles throughout the system are discussed as well, and these
exhibit qualitative differences from the linear chain, especially sublattice
character and a loss of particle-hole symmetry. 

\begin{figure}
{\centering \resizebox*{3.3in}{!}{\includegraphics*{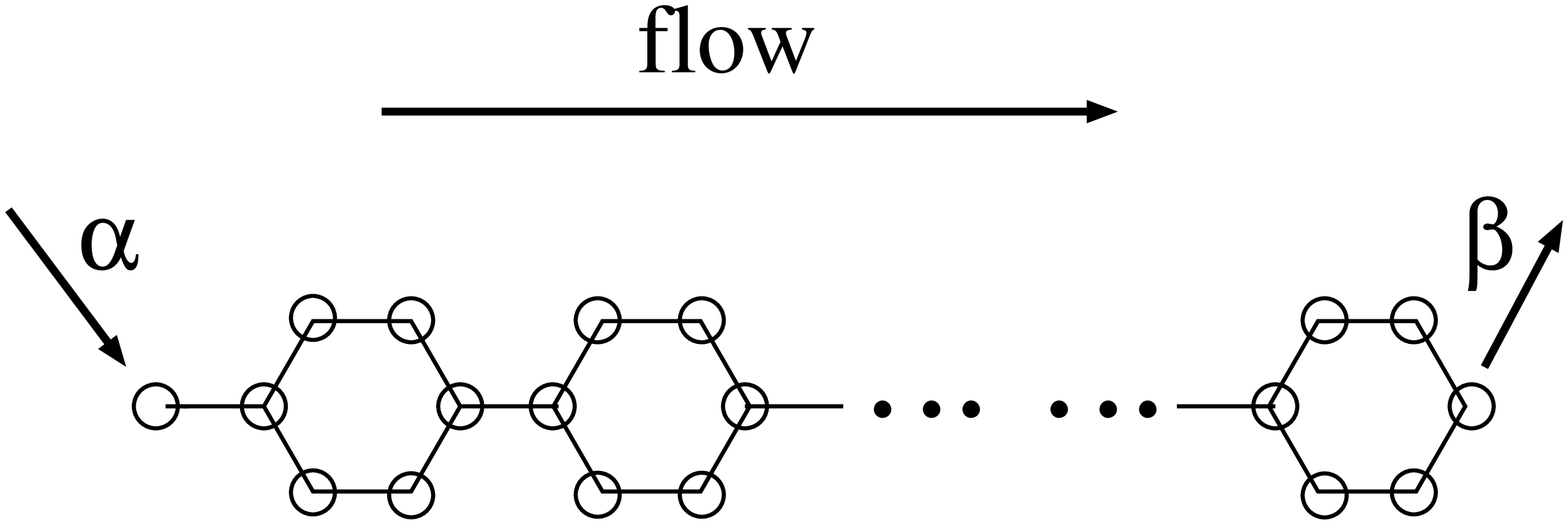}}}
\caption{
Necklace structure for TASEP with open boundary conditions at both ends, and corresponding 
injection and ejection rates $\alpha$ and $\beta$, respectively. Boundary conditions 
across direction of flow are free. 
} 
\label{fig:necklace}
\end{figure}

\begin{figure}
{\centering \resizebox*{2.0in}{!}{\includegraphics*{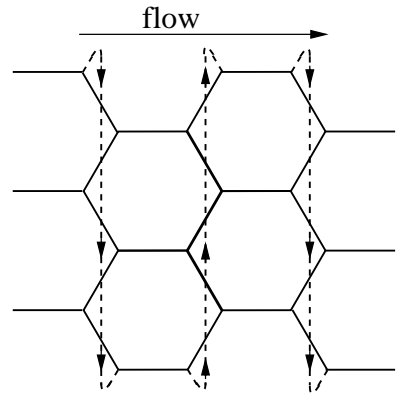}}}
\caption{
Planified section of a nanotube structure with $N_w=3$ hexagons round. The dashed lines
indicate the "wraparound" bonds which fulfil periodic boundary conditions
across the flow direction (for a nanoribbon, such bonds would be absent).
For clarity, bond directionalities are omitted, except for wraparound bonds.
} 
\label{fig:nnt}
\end{figure}

The most basic structure which, while departing as little as possible from the well-known 
strictly one-dimensional case, already displays 
sites with three-fold coordination,  is the necklace depicted in 
Fig.~\ref{fig:necklace}. Accounting for the direction of 
current flow, such sites can act either as "forking" points, or as "bottlenecks".
Boundary conditions perpendicular to the flow direction are free. For the case of
Fig.~\ref{fig:necklace} one has open boundary conditions at both ends. There, the 
externally-imposed parameters
are: the injection (attempt) rate $\alpha$ at the left end,
and the ejection rate $\beta$ at the right one.
Generalizations of the necklace are the cylinder-- (nanotube) or 
ribbon--like structures (see Fig.~\ref{fig:nnt}).
As seen in that Figure, the nanotube and ribbon
geometries considered here correspond, respectively, to zigzag (CNT) and 
armchair (CNR) configurations of the quasi-1D carbon allotropes~\cite{RMP}.
These configurations have no bonds orthogonal to the mean flow 
direction; thus they fall easily 
within the generalized TASEP description to be used, where each bond is 
to have a definite directionality, compatible with that of average flow.

All these structures are amenable to the mean field approach introduced 
and developed in Sections~\ref{subsec:theo-intro} and~\ref{subsec:theo-mf}. 
Sections~\ref{subsec:theo-bc} and~\ref{subsec:ext} concern boundary effects and 
extensions. Some special cases are highlighted in which exact solutions are
possible.

Numerical tests of the theory are given in Section~\ref{sec:num}.
In Section~\ref{subsec:intro-num} we describe the general
approach, pointing out details of the calculational method which
are expected to reflect properties of the actual transport process
in graphene-like samples.  
Section~\ref{subsec:neck-num} provides results for the necklace structure.   
Section~\ref{sec:nanotube} 
deals with honeycomb structures of arbitrary width with PBC
across the flow direction (nanotubes), and gives numerical results of pertinent
simulations. In Section~\ref{sec:ribbon} we consider honeycomb structures (ribbons) 
with free boundary conditions across the flow direction, and report results of numerical 
simulations.
In Section~\ref{sec:conc}, we summarize our results, and discuss 
the possible pertinence of the TASEP model results in the context of 
transport in  physical systems such as CNT, CNR, and quantum dot arrays. 
Concluding remarks are also presented there.  

\section{Theory}
\label{sec:theory}

\subsection{Introduction}
\label{subsec:theo-intro}
The emphasis here and throughout the paper is on steady-state properties of the 
TASEP on generalized geometries.
The microscopic variables, i.e., occupation probabilities $\tau_i$ for each site $i$,
satisfy a hierarchy of dynamic equations each relating $n-$ and $(n+1)-$body 
correlations.

In the steady state, the first of these becomes mean current conservation at any site.
Even here exact solution (requiring the whole hierarchy) is difficult, but can
be achieved in the simplest case of the linear chain with uniform bond 
rates~\cite{derr98,sch00,rbs01}. For this case mean field (factorization of correlations)
already gives an extremely useful account of steady-state properties, some of which,
like critical current, are exactly provided.

In what follows, the mean field procedure is extended to the new geometries.  
With uniform bias, equal average site occupations give (in mean field) equal currents on 
each bond. This gives a steady state for the chain. However, all the geometries 
considered here have "branchings" at sites with coordination number $z=3$, where
typically the division or merging of average current prevents equal site occupation
from giving a steady state. There are exceptions, e.g., where non-uniform bond rates
compensate. Except when this occurs, the steady states have a sublattice character.
The simplest of them have mean site occupations uniform on each of a number of
sublattices.

For analytic tractability we shall only consider cases where mean flow direction is 
parallel to one of the lattice directions,
and bond rates are independent of coordinate transverse to the flow direction.

For the chain, no sublattice division occurs but it is well known that in general the
mean-field site occupation profile has monotonic variations along the flow direction,
increasing for $J<J_c$ and decreasing for $J>J_c$, where $J_c$ is the critical
current dividing the two phases the profile characterizes. This result emerges
from a Mobius-type profile map with $J$-dependent coefficients which relates,
for specified $J$, the mean occupation of a site to that of the previous 
site~\cite{mukamel}.

In the generalized cases, the sublattice structure emerges directly from the
detailed form of the mean-field current conservation equations, in terms of
$J$ and all bond rates. Mobius maps for the profiles on each sublattice are given
by elimination of sites on other sublattices. By procedures similar to that
for the chain, the fixed points of the maps yield the special steady states
which are uniform on sublattices, as well as critical currents. Away from
the fixed points the maps give the spatially dependent generalizations, 
characteristic lengths etc. Various 
special characteristics for the chain are generalized, and the particle-hole 
exchange symmetry known for the chain typically disappears. 

\subsection{Mean field approach}
\label{subsec:theo-mf}

\begin{figure}
{\centering \resizebox*{3.0in}{!}{\includegraphics*{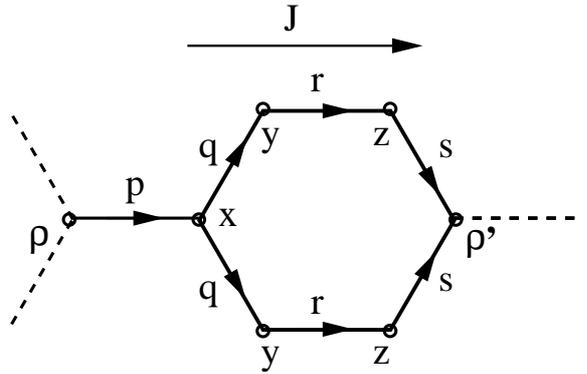}}}
\caption{
Bond rates $p$, $q$, $r$, $s$ and sublattice occupations $\rho$, $x$, $y$, 
$z$, $\rho^\prime$ for TASEP on necklace structure; see 
Eq.~(\protect{\ref{eq:js1}}).
} 
\label{fig:necksubl}
\end{figure}

We first consider the necklace. The mean field current across a bond with hopping rate
$p_{ij}$ going from site $i$ to site $j$ is $p_{ij}\,\langle \tau_i\rangle
\left(1-\langle\tau_j\rangle\right)$, where $\langle \tau_i\rangle$, 
$\langle \tau_j\rangle$ are the mean occupations of the two sites. 
The steady-state conservation equations for mean current $J$ are, for the
necklace section shown in Fig.~\ref{fig:necksubl}:
\begin{equation}
J=p\rho(1-x)=2qx(1-y)=2ry(1-z)=2sz(1-\rho^\prime)\ .
\label{eq:js1}
\end{equation}
These each relate site occupations $\rho$, $x$, $r$, $s$, and $\rho^\prime$ on
successive sublattices.

Eliminating site occupations between $\rho$ and $\rho^\prime$, i.e. on the
sublattices other than that which corresponds to $\rho$, $\rho^\prime$,
gives the relation for specified $J$:
\begin{equation}
\rho^\prime= \frac{a\rho - b}{c\rho-d}\ ,
\label{eq:map1}
\end{equation}
where
\begin{eqnarray}
a=4pqrs-2J\,\left[pqr+pqs+prs\right]+J^2\,pr \hskip0.4truecm\nonumber \\
b=2qJ\,\left[2rs-J(r+s)\right] \hskip 2.9truecm \nonumber \\
c=4pqrs-2Jps(q+r) \hskip 3truecm \nonumber\\
d=2qsJ\,\left[2r-J\,\right]\hskip 3.95truecm
\label{eq:abcd1}
\end{eqnarray}
The density profile maps for the other sublattices have the same form but with cyclically
interchanged rate variables. The map Eq.~(\ref{eq:map1}) is of Mobius form; the 
corresponding Mobius profile map for the TASEP chain~\cite{mukamel} has $d=0$, $a=c$. 
This simplification is related to a particle-hole symmetry, which is absent in the
general necklace, but is restored where the rates satisfy $a+d=c$ (needing $p=2s$,
$r=q$, see below).  

Iteration of the map for any sublattice gives that sublattice's 
density profile. 
Alternatively, one can use any one sublattice map, e.g. Eq.~(\ref{eq:map1})
with Eq.~(\ref{eq:abcd1}), together with Eq.~(\ref{eq:js1}), to give all details
(including relationships) of the sublattice density profiles. So, among other
things, all profiles are critical at the same $J_c$.

Assigning a site label $\ell$, increasing to the right, for each
sublattice the map Eq.~(\ref{eq:map1}), rewritten as $\rho_{\ell+1}=M(\rho_\ell)$,
gives the density profile $\{\rho_\ell\}$ for the "chosen" sublattice. 
The ansatz (see, for the chain, Refs.~\onlinecite{mukamel},~\onlinecite{rbs01})
\begin{equation}
\rho_\ell=A+B\tanh \theta_\ell\ ,\quad{\rm where}\quad \theta_{\ell+1}=
\theta_\ell+\phi\ ,
\label{eq:A23}
\end{equation}
is consistent with the map provided $\tanh \phi=cB/(cA-d)$ [$\,$from
decomposing $\tanh (\theta_\ell+\phi)\,$] and [$\,$to satisfy the remaining relations
for all $\theta_\ell\,$]
\begin{equation}
B(a+d)=2cAB\ ;\quad A(a+d)=b+c(A^2+B^2)\ .
\label{eq:A23b}
\end{equation}
In terms of $\ell_0$ such that $\theta_\ell=\phi\ell+\theta_0 \equiv 
\phi(\ell-\ell_0)$, one gets:
\begin{equation}
\rho_\ell=A+B\,\tanh \left\{\phi (\ell -\ell_0)\right\}\ ,
\label{eq:tanh}
\end{equation}
where
\begin{eqnarray}
A=\frac{a+d}{2c}\ ,\quad B=\frac{1}{2c}\sqrt{(a+d)^2-4bc}\ ,
\label{eq:AB}\\
\tanh \phi=\frac{1}{a-d}\sqrt{(a+d)^2-4bc}\ .
\label{eq:phi}
\end{eqnarray}

$A$, $B$, and $\phi$ are all dependent on $J$, since the coefficients
$a$, $b$, $c$, $d$ are. For a given set of bond rates, increasing $J$
can take it through a critical value $J_c$ at which the square root
vanishes and then becomes imaginary; then
\begin{equation}
\rho_\ell=A- |B|\,\tan \left\{|\phi| (\ell -\ell_0)\right\}\ ,\quad J>J_c
\label{eq:tan}
\end{equation}
while Eq.~(\ref{eq:tanh}) above applies with real $B$, $\phi$ for $J<J_c$.
This corresponds to a phase change, similarly to the TASEP chain. There,
and in the generalized systems being considered, $|\phi |$ is an inverse
characteristic length, which diverges at the (continuous) transition. For the
chain, but not in general, $A=1/2$, corresponding to the particle-hole symmetry
there, and absent for the generalizations.

As for the chain, the fixed points $\rho^\ast=\rho^>,\,\rho^<=A\pm B$ of the
controlling maps provide the special constant (sublattice) profiles, for $J<J_c$.
As $J \to J_c$, $\rho^>$ and $\rho^<$ come together, i.e. $B$ goes to zero,
as does the inverse length $|\phi |$, corresponding to criticality.

\begin{figure}
{\centering \resizebox*{1.8in}{!}{\includegraphics*{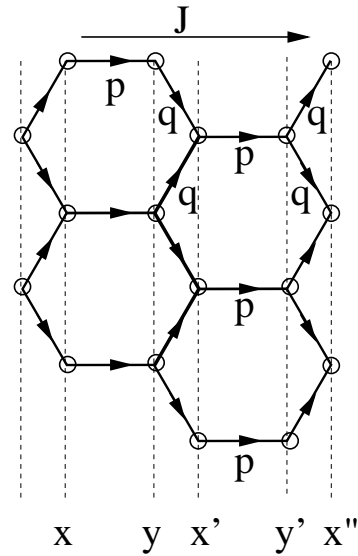}}}
\caption{
Bond rates $p$, $q$, and sublattice occupations  $x$, $y$, 
$x^\prime$, $y^\prime$, $x^{\prime\prime} \dots$ for TASEP on nanotube structure.
For clarity, wraparound and a few other bonds have been omitted [$\,$refer to 
Fig.~\ref{fig:nnt}$\,$]. See Eq.~(\protect{\ref{eq:js2}}).
} 
\label{fig:nntsubl}
\end{figure}

For the nanotube section  with the rates and mean site densities 
shown in Fig.~\ref{fig:nntsubl},
the steady-state current balance equations analogous to Eq.~(\ref{eq:js1}) are:
\begin{equation}
J=p\,x(1-y)=2qy(1-x^\prime)=p\,x^\prime(1-y^\prime)= \cdots
\label{eq:js2}
\end{equation}

The greater symmetry implies only two sublattices, 
and gives a simpler description than 
for the necklace. The consequent sublattice Mobius map for the $x$-sublattice is:
\begin{equation}
x^\prime\ (=x_{\ell+1})\ =\frac{ax-b}{cx-d}=M(x=x_\ell)\ ,
\label{eq:map2}
\end{equation}
with
\begin{equation}
a=p(2q-J)\ ;\quad b=d=2qJ\ ;\quad c=2pq\ .
\label{eq:abcd2}
\end{equation}
Eqs.~(\ref{eq:tanh})--(\ref{eq:tan}) also apply here, with  $a$, $b$, $c$, $d$
now given by Eq.~(\ref{eq:abcd2}).  
The map for the $y$-sublattice is similar, but with $2q$ and $p$ interchanged
in the expressions for $a$, $b$, $c$, $d$. This implies that for the special case
\begin{equation}
2q=p\qquad\qquad{\rm (nanotube)}
\label{eq:rates1}
\end{equation} 
the two sublattice maps are identical. In this case, for adjacent sites in the full
sequence $x$, $y$, $x^\prime$, $y^\prime$, $x^{\prime\prime}$, $y^{\prime\prime}$, 
$\cdots$ the map is always the same as for the linear chain TASEP. In this sense,  
for $2q=p$ sublattices are irrelevant and the mean-feld steady state for the nanotube 
is equivalent to that on the linear chain; the particle-hole symmetry is also recovered.

There is also a special case for the necklace, with the rates 
(see Fig.~\ref{fig:necksubl})
\begin{equation}
q=r=s= \frac{p}{2}\qquad\qquad{\rm (necklace)}
\label{eq:rates2}
\end{equation} 
for which it is easy to check that the sublattice profile maps are the same as the
fourth-iterated linear chain map. So, as for the nanotube with $2q=p$ the sublattice
density profiles are the same as for the chain, except for spatial and rate 
rescalings.

Eqs.~(\ref{eq:tanh}),~(\ref{eq:AB}),~(\ref{eq:phi}), and~(\ref{eq:tan}) apply
equally well to general necklace and nanotube, as do their detailed consequences 
such as critical current.
Since  $J_c$ is where $B$ and $\phi$ of Eqs.~(\ref{eq:AB}) 
and~(\ref{eq:phi}) vanish, $J_c$ is given by solving  the relation
\begin{equation}
(a+d)^2=4bc
\label{eq:rel}
\end{equation}
between the $J$-dependent coefficients, given respectively by Eq.~(\ref{eq:abcd1})
for the necklace, and Eq.~(\ref{eq:abcd2}) for the nanotube.

For the necklace this leads to a quartic equation for $J_c$, which factorizes
for the special case of Eq.~(\ref{eq:rates2}), yielding $J_c=p/4$ as in the
equivalent linear chain. Two additional results are
worth recording, for comparison with numerical tests of the theory:
\begin{eqnarray}
p=r=s=2q:\quad J_c=\frac{p}{3}\ ;\hskip1.5truecm
\label{eq:necksp1}\\
p=q=r=s:\quad \frac{J_c}{p}=0.40277\dots\ .
\label{eq:necksp2}
\end{eqnarray} 

For the nanotube the equation for $J_c$ is quadratic for general rates,
resulting in 
\begin{equation}
J_c= \mu_2^{-2}\,\left[\mu_0-\sqrt{\mu_0^2-\mu_1^2\mu_2^2}\,\right]\ ,
\label{eq:jc1}
\end{equation}     
where $\mu_0=2pq(2q+p)$, $\mu_1=2pq$, $\mu_2=2q-p$, as long as $\mu_2 \neq 0$;
when $p=2q$ as in Eq.~(\ref{eq:rates1}), $J_c=p/4$; this is the case equivalent
to the chain. Among other special cases needed later for comparison with simulations
is $p=q$, where
\begin{equation}
J_c/p=2[3-\sqrt{8}]\qquad({\rm nanotube},\quad p=q)\ .
\label{eq:jcntpq1}
\end{equation}
The above analysis can also provide the characteristic length $\xi \equiv
1/|\phi|$ (kink width, etc). For $J$ near $J_c$, $\xi$ is found to diverge like
$|J-J_c|^{-1/2}$, in general, as for the linear chain. This can be used in the
scaling analysis of finite size corrections for the current etc, but not too close
to the critical point where fluctuation effects absent from mean field theory are
expected to dominate, changing the above exponent.

\subsection{Boundary effects}
\label{subsec:theo-bc}

Boundary effects are strongly affected by the sublattice distinctions required
in the generalized geometries. Here the sublattice $\tanh$- or $\tan$- 
density profiles given, 
respectively, in Eqs.~(\ref{eq:tanh}) and~(\ref{eq:tan}) are qualitatively
similar to the linear 
chain case. The current conservation equations Eqs.~(\ref{eq:js1}),~(\ref{eq:js2})
show that here, for example a $\tanh$ solution on one sublattice requires one
on the other sublattices, so all sublattices are in the same (low- or
maximal-) current phase; also, for $J<J_c$ where kinks are present in the
chain density profile, the same equations imply that the kinks on different
sublattices are in neighboring positions. 
So for example, in an open geometry, a kink near left, or right, or 
neither boundary on one sublattice implies the same on the other sublattices,
so different sublattices also share the same low- or high-density, or
coexistence character.

But the details depend crucially on which sublattices the boundary sites 
sit on, and in particular whether they are on the same sublattice. The 
same is true with PBC. In the following, where we
examine the influence of boundary conditions along the flow direction,
we consider only the case of boundary sites on the same sublattice
[$\,$which we take as the "chosen" sublattice in 
Section~\ref{subsec:theo-mf}, see Eqs.~(\ref{eq:map1}) 
and~(\ref{eq:map2})$\,$]. This applies when an integer number $N$ of
basic units (one bond attached to the left of a full hexagon) span
the system. Any required generalization could be readily made using
the relationships between sublattices provided by the current
conservation equations (not done here).

\subsubsection{PBC}
\label{theo:pbc}

For PBC of the special type just defined, as in the linear chain, the sublattice 
steady-state density profiles are flat. 
We focus on the mean-field "fundamental relation" between $J$ and  
$\langle \rho\rangle$, 
where the latter is the (sublattice-averaged) global mean density.
This is found by obtaining the two flat density profiles for a chosen sublattice,
then using the current conservation equations to obtain the corresponding
flat density profiles for the other sublattices, and combining them
with the correct weights to find $\langle \rho\rangle$, for the 
given $J$. 

For the nanotube this reduces to
\begin{equation}
\langle \rho\rangle -\frac{1}{2}=\pm B(J)=\pm \frac{1}{2c}\sqrt{(a+d)^2-4bc}\ ,
\label{eq:rho}
\end{equation}
with $a$, $b$, $c$, $d$ given by Eq.~(\ref{eq:abcd2}). Inversion gives 
the fundamental
relation, which is always quadratic sufficiently close to the maximum $J_c$.
Remarkably, for any $p$, $q$ the maximum occurs at $\langle 
\rho\rangle=1/2$, same as for the linear chain.
For the special case $p=2q$ of Eq.~(\ref{eq:rates1}) the full nanotube
result reduces exactly everywhere to
\begin{equation}
J = \frac{p}{4}-p\left(\langle\rho\rangle - \frac{1}{2}\right)^2\ ,
\label{eq:ntpeq2q}
\end{equation}
as for the chain.

For the necklace the current-density relation is also quadratic close to $J_c$:
\begin{equation}
J_c-J \propto \left(\langle \rho \rangle -\rho_{\rm max}\right)^2\ .
\label{eq:f.rel}
\end{equation}
However, $\rho_{\rm max} \neq 1/2$, except in the special cases with $s=q$
[$\,$including that of Eq.~(\ref{eq:rates2}) with sublattice equivalence to the
linear chain$\,$] where a symmetry argument applies, based on particle - hole 
duality under flow reversal. 

In the special case of Eq.~(\ref{eq:necksp1}),
\begin{equation}
J= J_c-p\,\frac{32}{27}\left(\langle \rho \rangle -\frac{17}{36}\right)^2 + \dots
\label{eq:nlspec}
\end{equation}
for $J$ near $J_c=p/3$~.

\subsubsection{Open BC}
\label{theo:obc}

We next consider open boundary conditions, again with boundary sites
on the same "chosen" sublattice.

In the mean field approach, the injection/ejection processes at the left/right ends, 
with respective attempt rates $\alpha$ and $\beta$, generalize the current conservation
conditions expressed in Eqs.~(\ref{eq:js1}) and~(\ref{eq:js2}) by the extra
equations:
\begin{equation}
J=\alpha\left(1-\rho_{\ell=0}\right)=\beta\rho_{\ell=N}\ .
\label{eq:bdy1}
\end{equation}
For given internal bond rates these in principle give the, so far free,
variables $J$ and $\ell_0$, and hence everything in terms of
$\alpha$, $\beta$. At the critical condition, 
where $B$ and $\phi$ of Eqs.~(\ref{eq:AB}) 
and~(\ref{eq:phi}) vanish, the extra equations give the critical point
in the $(\alpha,\beta)$ plane as
\begin{equation}
(\alpha_c,\beta_c)=\left(\frac{J_c}{1-A(J_c)}\,,\,\frac{J_c}{A(J_c)}\right)\ ,
\label{eq:acbc1}
\end{equation} 
generally without the symmetry of the critical point $(\alpha_c,\beta_c)=
(p/2,p/2)$ for the chain, where $A=1/2$, $J_c=p/4$. E.g., for the
nanotube:
\begin{equation}
(\alpha_c,\beta_c)=\left(\frac{2\mu_1\,J_c}{\mu_1-\mu_2\,J_c}\,,\,
\frac{2\mu_1\,J_c}{\mu_1+\mu_2\,J_c}\right)\ ,
\label{eq:acbc2}
\end{equation}
in terms of the variables defined in connection with Eq.~(\ref{eq:jc1}),
while for the necklace with $p=r=s=2q$,
\begin{equation}
(\alpha_c,\beta_c)=\left(p,\frac{p}{2}\right)\ .
\label{eq:acbc3}
\end{equation} 
By considering sublattice kinks near the system's boundaries, it can be
shown that in mean-field theory the phase boundaries are vertical and
horizontal lines in the $\alpha-\beta$ plane through  and outwards
from the critical point. 

With $(\alpha,\beta)$ sufficiently below $(\alpha_c,\beta_c)$ the equations are
consistent with $J<J_c$ and kink width $\phi^{-1}$ not very large. 
Then the
constraint equations, Eq.~(\ref{eq:bdy1}), can be consistent with having
the sublattice kinks away from the boundaries, so that the relations of
$\alpha$, $\beta$ to profile values $A \pm B$ are:
\begin{equation}
\alpha\,\left[ 1-\left(A(J)-B(J)\right)\right]= \beta\,\left[A(J)+B(J)\right]=
J<J_c\ .
\label{eq:bdy2}
\end{equation} 
For given rates this is a parametric equation in $J$ for a curve 
(the coexistence line) in the $\alpha-\beta$ plane. 
For the nanotube this becomes
\begin{equation}
2pq(\alpha-\beta)=(2q-p)\alpha\beta\ .
\label{eq:bdy2b}
\end{equation}
For both necklace and nanotube it can be easily 
established that the coexistence line joins the origin to the 
critical point, and though it is in general not straight its slope at 
the origin is always unity.

\subsection{Extensions}
\label{subsec:ext}

In the preceding mean-field discussion, the simplest situation has been
where profiles are flat on sublattices. For the chain this is known to
be an exact property under special conditions where correlation functions 
factorize; then size dependences disappear~\cite{derr93}. Here we consider 
this possibility for necklaces and nanotubes with open boundary conditions.

We start by assuming factorization in the sense that occupations 
$\tau_i$ are the same on all sites of a sublattice but different between 
sublattices. This is consistent with exact average-current conservation 
(the first member of the hierarchy of exact steady state correlation
function equations). Since no occupation
variable occurs squared in those equations, it is also consistent with
our generalized mean field approximation, including the Mobius 
maps and their
consequences as spelt out in Section~\ref{subsec:theo-mf}, 
but only under conditions where those give a constant profile
on each sublattice, i.e., fixed points $\rho^\ast$.
 Adding the injection/ejection constraints, 
Eq.~(\ref{eq:bdy1}), the flat fixed point profiles will not in general be
consistent with the latter, unless
\begin{equation}
\alpha\,\left[1-\rho^\ast(J)\right] =J =\beta\,\rho^\ast(J)\ .
\label{eq:bdy3}
\end{equation}

So a necessary condition for the factorization to give an exact
solution is that ($\alpha$, $\beta$) lies on the line in the
 $(\alpha ,\beta)$ plane whose parametric equation is Eq.~(\ref{eq:bdy3}).
The remaining condition for sufficiency is that factorization is
consistent with all the other members of the hierarchy of internal
correlation function equations (not proven here).

For the general nanotube, elimination of $J$ from Eq.~(\ref{eq:bdy3}) gives the
"factorization line" as
\begin{equation}
\frac{\alpha}{2q} + \frac{\beta}{p} =1\ .
\label{eq:line}
\end{equation}
This line goes through the critical point $(\alpha_c,\beta_c)$ of
Eq.~(\ref{eq:acbc2}), and becomes the same as for the chain if $2q=p$
[$\,$see Eq.~(\ref{eq:rates1})$\,$].

Quadratic $J$-dependences in the map coefficients for the necklace complicate the 
analysis, but the line (now curved) again goes through 
$(\alpha_c,\beta_c)$ of
Eq.~(\ref{eq:acbc2}); the special case $q=r=s=p/2$ again becomes that for the chain
[$\,$see Eq.~(\ref{eq:rates2})$\,$].
 
In Section~\ref{sec:num} we provide numerical checks of selected predictions
of the mean-field theory just described, namely steady-state currents and their
dependence on average particle density (for PBC) or 
on injection/ejection attempt rates
(for systems with open boundaries), for both necklaces (Section~\ref{subsec:neck-num})
and nanotubes (Section~\ref{sec:nanotube}), all for assorted bond rate combinations
of interest. 
For nanoribbons the translational symmetry perpendicular
to flow direction (crucial in the reduction of the number of sublattices for the
nanotube, see Eq.~(\ref{eq:js2}) and Fig.~\ref{fig:nntsubl}) is lost with
free boundary conditions at the edges. Thus, in this case we restricted ourselves
to the numerical simulations described in Sec.~\ref{sec:ribbon}~.

Finally, we performed some numerical tests of factorization for nanotubes with
open boundary conditions [$\,$see Eqs.~(\ref{eq:bdy3}) and~(\ref{eq:line}) 
above$\,$]. They are briefly reported at the end of Sec.~\ref{sec:conc}. 

\section{Numerics}
\label{sec:num}

\subsection{Introduction}
\label{subsec:intro-num}
For simplicity we consider structures with an integer number $N_r$ of elementary cells (one bond 
attached to the left of a full hexagon) along the mean flow direction.

Adapting the procedures used for the $(1+1)-$ dimensional TASEP, an elementary time step consists of
$N_b$ sequential bond update attempts, each of these according to the following rules: (1) select
a bond at random, say, bond $\ell$; (2) if the chosen bond has an occupied site to its left and an 
empty site to its right, then (3) move the particle across it with probability (bond rate)  $p_\ell$.
If the injection or ejection bond is chosen, step (2) is suitably modified to account for the 
particle reservoir (the corresponding bond rate being, respectively,  $\alpha$ or $\beta$). 

One can equally well update sites instead (via $N_s$ random sequential site choices). 
Once a site is picked, (i) if the site is a "forking" one, either of the two bonds to its right is
randomly selected [$\,$with probability $1/2$, i.e., no transverse bias is allowed$\,$], and then
steps (2) and (3) above are followed; (ii) for open boundary conditions, if the site is the injection 
(ejection) one, then if it is unoccupied (occupied), a particle is injected into (ejected out of) it
with probability $\alpha$ ($\beta$); (iii) otherwise, steps (2) and (3) above are followed right away. 

It is easily seen that on average a total of $N_b$ update attempts 
will take place, in the course of a unit 
time step as defined above, for either bond or site update.  
In the strictly one-dimensional TASEP, bond- and site update are entirely equivalent.
However, for full equivalence between the two methods in the present case, it must be noted that 
randomly selecting (with $1/2$ probability) which bond to probe, when starting from a "forking" 
site, effectively {\em halves} the following bonds's rates. For all geometries investigated 
here we ran simulations using both bond and site update. In all cases for which the effective
bond rates (i.e. taking into account the effect just described for site update) coincided, the 
results given by both methods were indistinguishable within error bars.

Both site and bond update may be relevant in physical applications.  We defer a
discussion of the  potential relationship of each update method to specific features of 
graphene-like  structures to Sec.~\ref{sec:conc}. 

Here we evaluate the steady-state current $J$ as the time- and ensemble-averaged number of particles (per unit time) 
which (a) enter  the system [$\,$for open ends$\,$], or (b) cross any single bond connecting adjacent 
hexagons [$\,$for PBC$\,]$. For systems with $N_e>1$ "entry" bonds, such as the
nanotubes and ribbons considered, respectively, in Sections~\ref{sec:nanotube} and~\ref{sec:ribbon}, one
has to divide further by $N_e$, to provide proper comparison with the strictly one-dimensional case.     
Starting from a spatially random configuration of 
occupied and empty sites, we usually waited $n_{\rm in}=10,000$ time steps for steady-state flow to be fully
established, to ensure that our measurements were free from startup effects. After that, we
collected steady-state current samples (typically for $N_{\rm sam}=10^{\,6}$
consecutive unit time steps). The accuracy of results was estimated by evaluating
the root-mean-square (RMS) deviation  among $N_{\rm set}$ 
independent sets of $N_{\rm sam}$ steady-state samples each. As is well known~\cite{rbsdq12}, such RMS deviations
are essentially independent of $N_{\rm set}$
as long as $N_{\rm set}$ is not too small, and vary as $N_{\rm sam}^{-1/2}$.
We generally took $N_{\rm set}=10$.

\subsection{The Necklace Structure}
\label{subsec:neck-num}
The structures considered here have $N_s^P=6N_r$ sites and 
$N_b^P=7N_r$ bonds (for PBC), or $N_s^O=6N_r+1$ sites [$\,$recall the 
extra site on the right, connecting to the ejection bond, see 
Fig.~\ref{fig:necklace}$\,$] and $N_b^O=7N_r+2$ 
bonds (counting the injection and ejection bonds, for open boundary conditions. 

We first check the mean-field prediction, see Eq.~(\ref{eq:rates2}),
that the steady-state current on a system where all bond rates on the
hexagons are $1/2$, and those on bonds between hexagons are unity, is the same as on a strictly one-dimensional
arrangement.  Using site updating procedures, we fixed the nominal bond rates for 
links on a hexagon immediately following a "forking" site to be unitary,
so they would effectively be halved. We also ran simulations using bond update, in which 
case all hexagon bonds were set to $p=1/2$ from the start, with the same results 
(within error bars) as those from site update.

For systems with PBC, the current on a strictly one-dimensional lattice with $N$ sites and
$M$ particles (average density $\rho=M/N$), and unit bond rates, is~\cite{mukamel}
\begin{equation}
J =\rho\,(1-\rho)\,\frac{N}{N-1}\quad\ (d=1,\ {\rm PBC})\ .  
\label{eq:j1dpbc}
\end{equation}
Table~\ref{t1} illustrates the excellent agreement found between theoretical predictions 
and numerical simulations for PBC. Remarkably, the identification between 
necklace- and chain current goes as far as finite-size effects: systems 
of either type with the same number of sites obey Eq.~(\ref{eq:j1dpbc}) 
equally. This is consistent with exact factorizability needing no conditions
like Eq.~(\ref{eq:bdy3}) in the case of PBC.

\begin{table}
\caption{\label{t1}
For systems with PBC, $N_s$ sites, and $\langle\rho\rangle$ as specified, $J_{\rm num}$
is current through necklace (with $N_r=N_s/6$ rings, effective 
bond rates $p=1/2$ on hexagons, $p=1$ on bonds between hexagons), as given  
by numerical simulations with $N_{\rm sam}=10^{\,6}$, $N_{\rm set}=10$ (see text);
$J_{\rm 1d}$ is current through one-dimensional system with $N=N_s$ , given by
 Eq.~(\protect{\ref{eq:j1dpbc}})~.
}
\vskip 0.2cm
\begin{ruledtabular}
\begin{tabular}{@{}ccc}
$\ \ N_s$ & $J_{\rm num}$  & $J_{\rm 1d}$    \\
\hline\noalign{\smallskip}
\multicolumn{3}{c}{$\langle\rho\rangle=1/2$}\\
\hline\noalign{\smallskip}
$\ \ 24$ & $0.26081(13)$     &  $0.260867\dots$ \\
$\ \ 36$ & $0.25713(18)$     &  $0.257143\dots$ \\
$\ \ 48$ & $0.25533(11)$     &  $0.255319\dots$ \\
$\ \ 60$ & $0.25426(11)$     &  $0.254237\dots$ \\
$\ \ 72$ & $0.25353(14)$     &  $0.253521\dots$ \\
$\ \ 84$ & $0.25297(10)$     &  $0.253012\dots$ \\
$\ \ 96$ & $0.25261(7)\ \,$  &  $0.252632\dots$ \\
\hline\noalign{\smallskip}
\multicolumn{3}{c}{$\langle\rho\rangle=1/4$}\\
\hline\noalign{\smallskip}
$\ \ 24$ & $0.19563(15)$     &  $0.195652\dots$ \\
$\ \ 36$ & $0.19280(7)\ \,$  &  $0.192857\dots$ \\
$\ \ 48$ & $0.19149(11)$     &  $0.191489\dots$ \\
$\ \ 60$ & $0.19066(9)\ \,$  &  $0.190678\dots$ \\
$\ \ 72$ & $0.19015(9)\ \,$  &  $0.190141\dots$ \\
$\ \ 84$ & $0.18971(5)\ \,$  &  $0.189759\dots$ \\
$\ \ 96$ & $0.18945(12)$     &  $0.189474\dots$ \\
\end{tabular}
\end{ruledtabular}
\end{table}

Next we give results for systems with open boundary conditions, 
also for the special rates of Eq.~(\ref{eq:rates2}), at selected locations on the 
$\alpha$--$\beta$ phase diagram; see Table~\ref{t2}. Agreement with mean-field theory 
(including finite-size effects, or their absence) is very good at
$(\alpha,\beta)=(1/2,1/2)$, as well as at $(1/4,1/4)$ [$\,$the latter point corresponding to the coexistence
line between high-and low-density phases in the one-dimensional TASEP$\,$]. For  
$(\alpha,\beta)=(1,1)$, deep within the maximal-current phase of the one-dimensional problem, small discrepancies
are present for small systems; however, they tend to vanish as $N_s$ increases.
\begin{table}
\caption{\label{t2}
For systems with open boundary conditions, $N_s$ sites, and $(\alpha,\beta)$ as specified, $J_{\rm num}$
is current through necklace (with $N_r=(N_s-1)/6$ rings, effective 
bond rates $p=1/2$ on hexagons, $p=1$ on bonds between hexagons), as given  
by numerical simulations with $N_{\rm sam}=10^{\,6}$, $N_{\rm set}=10$ (see text);
$J_{\rm 1d}$ is current through one-dimensional system, see e.g.
 Refs.~\protect{\onlinecite{derr93,rbsdq12}}~.
}
\vskip 0.2cm
\begin{ruledtabular}
\begin{tabular}{@{}ccc}
$\ \ N_s$ & $J_{\rm num}$  & $J_{\rm 1d}$    \\
\hline\noalign{\smallskip}
\multicolumn{3}{c}{$(\alpha,\beta)=(1/2,1/2)$}\\
\hline\noalign{\smallskip}
$\ \ 13$ & $0.24993(26)$     &  $1/4$ \\
$\ \ 31$ & $0.24996(15)$     &  $1/4$ \\
$\ \ 61$ & $0.24998(10)$     &  $1/4$ \\
$\ \ 301$ & $0.24999(9)\ \,$ &  $1/4$ \\
\hline\noalign{\smallskip}
\multicolumn{3}{c}{$(\alpha,\beta)=(1/4,1/4)$}\\
\hline\noalign{\smallskip}
$\ \ 13$ & $0.17479(24)$     &  $0.175399\dots$ \\
$\ \ 31$ & $0.18175(25)$     &  $0.181903\dots$ \\
$\ \ 61$ & $0.18448(27)$     &  $0.184547\dots$ \\
$\ \ 301$ & $0.18688(25)$  &    $0.186882\dots$ \\
\hline\noalign{\smallskip}
\multicolumn{3}{c}{$(\alpha,\beta)=(1,1)$}\\
\hline\noalign{\smallskip}
$\ \ 13$ & $0.28303(18)$     &  $0.277777\dots $ \\
$\ \ 31$ & $0.26324(13)$     &  $0.261905\dots$ \\
$\ \ 61$ & $0.25543(7)\ \,$  &  $0.256098\dots$ \\
$\ \ 301$ & $0.25125(10)$    &  $0.251244\dots$ \\
\end{tabular}
\end{ruledtabular}
\end{table}

We now turn to combinations of bond rates for which mean-field solutions 
are less simple, but
which are plausible in terms of potentially describing electronic transport on a 
graphene-like structure. 
Assuming the simplest case of lattice homogeneity, we consider uniform rates $p=1$ 
for all bonds.
Except where otherwise noted, we use site update procedures in the simulations 
described here; thus, as explained above, the
bonds immediately following a "forking" site have their effective rates halved.

\begin{figure}
{\centering \resizebox*{3.3in}{!}{\includegraphics*{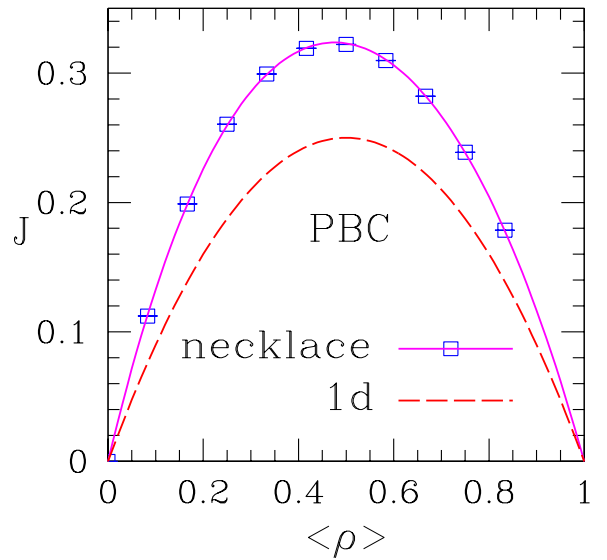}}}
\caption{(Color online) 
Current-density relationship for necklace structure with PBC, all nominal 
bond rates $p=1$, site update. Points correspond to simulations;
the solid curve is a fourth-degree polynomial fit to the data. The long-dashed curve is for the one-dimensional
TASEP with PBC.}
\label{fig:jrhonl}
\end{figure}

For the necklace with PBC we calculated steady-state currents for $\langle\rho\rangle=m/12$, 
$m=1,2, \cdots 11$.
For each density we considered rings with $N_r=4,6, \cdots, 16$ elementary cells.
The respective sequences behave smoothly against $N_r^{-1}$, and were extrapolated to $N_r^{-1} \to 0$ by 
fits to quadratic polynomials. Final results are shown in Fig.~\ref{fig:jrhonl}. 
As predicted in Sec.~\ref{theo:pbc}, for this case in which $q 
\neq s$ the particle-hole symmetry  is lost. The maximum of the
adjusted curve is at $(\langle\rho\rangle,J)=(0.475(3),0.3234(2))$~. The 
parabolic shape near the maximum, predicted in Eq.~(\ref{eq:nlspec}), is verified,
and the numerical value given there for $p=1$, namely $\rho_c=17/36=0.4722 \dots$, is 
within error bars; however, the predicted $J_c=1/3$ appears to overshoot the numerical
result by some $3\%$.

The values of $\rho_c$, $J_c$ given above are to be compared also with the
maximal current for the one-dimensional TASEP with PBC and unit bond rates, namely $J=1/4$ at $\rho=1/2$,
see Eq.~(\ref{eq:j1dpbc})~. 

\begin{figure}
{\centering \resizebox*{3.3in}{!}{\includegraphics*{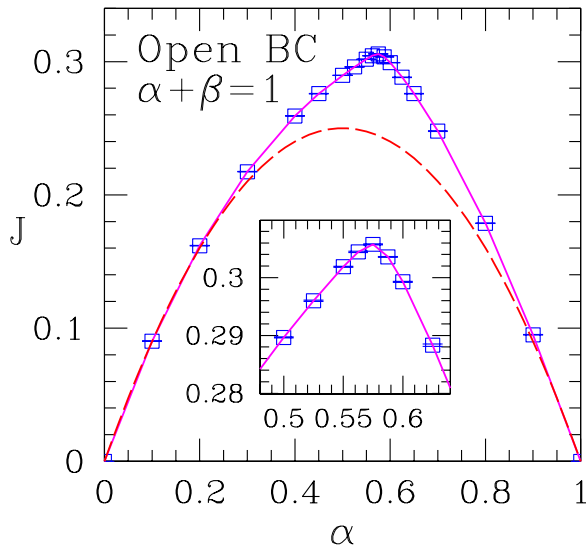}}}
\caption{(Color online) 
Current against injection rate $\alpha$ for necklace structure with open boundaries, 
along $\alpha+\beta=1$, all nominal bond rates $p=1$, site update. 
Points correspond to simulations. The long-dashed curve is for the one-dimensional
TASEP on the line $\alpha+\beta=1$. Inset: close-up view of peak region. Same axis labels as
main figure. }
\label{fig:jabnl}
\end{figure}

For the necklace with open boundary conditions, and nominal rates $p=1$ for 
all internal bonds [$\,$i.e. except for
the injection and ejection bonds at the extremes, with their characteristic rates $\alpha$ and $\beta\,$],
we first report results on the $\alpha+\beta=1$ line. System sizes were the same as for PBC.
The current $J$, parametrized by $\alpha$,
is shown in Fig.~\ref{fig:jabnl}. For the one-dimensional TASEP, the steady-state
current is $J=\alpha\beta$ on this line, and is size-independent~\cite{derr93}. Here we found  
little size dependence for both $\alpha \lesssim 0.4$ and $\alpha \gtrsim 0.7$. Around the peak shown
in Fig.~\ref{fig:jabnl}, the current distinctly increases with system size, thus we 
resorted to linear or quadratic fits against $N_r^{-1}$ to produce extrapolated values.
Comparison with the one-dimensional TASEP would suggest that an increase in $J$ with
system size indicates proximity to a coexistence line between low- and high-density phases, see the entries
for $(\alpha,\beta)=(1/4,1/4)$ in Table~\ref{t2}.
In contrast to the case of PBC, we could not produce a single, smooth fitting function for the
$J$ vs. $\alpha$ relationship over the full range $0 < \alpha < 1$, mainly because of the 
sharply 
asymmetric peak. We estimate the largest current along $\alpha+\beta=1$ to be $J=0.3058(2)$
at $\alpha=0.575(3)$.

\begin{figure}
{\centering \resizebox*{3.3in}{!}{\includegraphics*{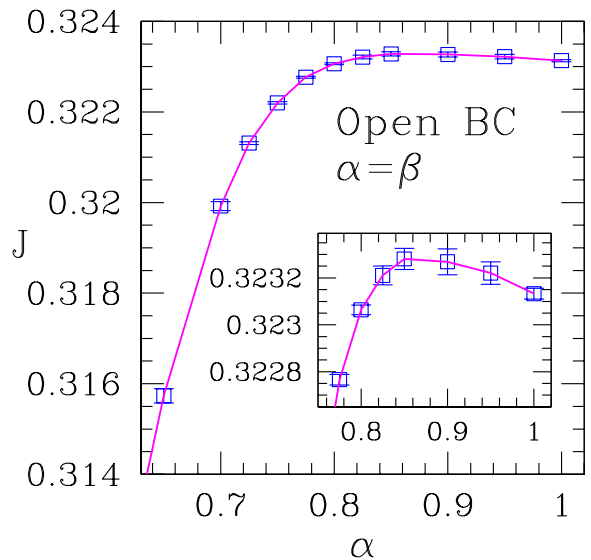}}}
\caption{(Color online) 
Current against injection rate $\alpha$ for necklace structure with open boundaries, 
along $\alpha=\beta$, all nominal bond rates $p=1$, site update. 
Points correspond to simulations. 
Inset: close-up view of region close to $\alpha=1$. Same axis labels as
main figure. }
\label{fig:jaeqbnl}
\end{figure}

Relying once more on analogies with the one-dimensional TASEP, we examined the region close to $(\alpha,\beta)=(1,1)$
in order to probe the extent of a hypothetical maximal-current phase. Fig.~\ref{fig:jaeqbnl} shows the extrapolated
($N_r \to \infty$) currents along $\alpha=\beta$, for $0.65 \leq \alpha \leq 1$. System sizes used were the same as
for PBC, except that for $\alpha>0.7$ we went up to $N_r=50$. In the latter region, improved accuracy was
necessary in order to distinguish between very similar values (see especially the inset of Fig.~\ref{fig:jaeqbnl}). 

Taking account of the error bars for individual results, our tentative conclusion is that the current 
indeed stabilizes at $J_{\rm max}=0.3232(1)$,  and that
the section of the $\alpha=\beta$ line for $\alpha \gtrsim 0.825$ is within the maximal-current phase.

The estimate just found for the maximal current is consistent within error
bars with the corresponding one for PBC, also with site update and same bond rates, 
namely  $0.3234(2)$. This is to be expected, since regardless of boundary conditions 
$J_c$ is achieved with density profiles $\rho_c= \rho^> =\rho^<$; 
for open boundary conditions this imposes additional constraints on $\alpha$, 
$\beta$ [$\,$see Eq.~(\ref{eq:bdy3})$\,$]  while PBC are 
automatically consistent with  flat profiles. 

We also checked the prediction of Eq.~(\ref{eq:necksp2}) for the critical current
on the necklace with all effective rates equal to unity, by using bond update procedures.
For PBC with this particular set of rates, mean field theory
predicts (see Sec.~\ref{theo:pbc}) that the $J-\langle \rho \rangle$ curve 
is symmetric about 
$\langle \rho \rangle=1/2$, thus restoring particle-hole symmetry.
A scan through various average densities, similar to that shown in 
Fig.~\ref{fig:jrhonl}, indeed resulted in a symmetric curve; however,
we found $J_c=0.3958(4)$, just under $2\%$ below the mean field prediction.
With open boundary conditions we scanned the region
of the $(\alpha,\beta)$ plane close to $(\alpha,\beta)=(1,1)$ and found
a picture qualitatively similar to the one exhibited in Fig.~\ref{fig:jaeqbnl}. 
From that we estimate $J_c=0.395(1)$, in good agreement with the PBC result.

\subsection{Nanotubes} 
\label{sec:nanotube}

We consider strips of a two-dimensional honeycomb lattice with the same orientation, 
relative to particle flow direction, as the necklace, and with periodic boundary conditions
across the flow direction (recall Figs.~\ref{fig:nnt} and~\ref{fig:nntsubl}). 
Such "nanotubes" can be seen as $N_w$ parallel necklaces, with
adjacent necklaces sharing edges parallel to the flow direction, as well as the corresponding
sites. The total number of sites is thus $N_s^P=N_w \times 4N_r$ for PBC, 
or $N_s^O=N_w \times (4N_r+1)$ for open boundary conditions at the ends.
As remarked in 
Sec.~\ref{subsec:neck-num}, the normalized current $J$ in this case is the (average) 
total number of particles moving through a fixed cross-section of the system, per unit time,
divided by $N_w$.

According to Eqs.~(\ref{eq:rates1}) and~(\ref{eq:ntpeq2q}),
for the case where all bonds parallel to the flow direction have rates
$p$, and all others have $p/2$, the current is the same as on a one-dimensional 
lattice with all bond rates equal to $p$. For the nanotube, such (effective) rates 
correspond to the physically plausible assumption of equal nominal rates $p=1$ on all 
bonds, together with the use of site update procedures.

We first examine a toroidal geometry, i.e., one with PBC in both directions. 
The finite-length effects on the one-dimensional lattice, which come via the
$N$-dependent factor in
Eq.~(\ref{eq:j1dpbc}), here correspond to the total number of sites on the nanotube, i.e. 
$N=N_w \times 4N_r$, independent of the aspect ratio $A \equiv N_w/4N_r$. 
This is illustrated by the results in Table~\ref{t3}.
\begin{table}
\caption{\label{t3}
For systems with $N_s$ sites, and $\langle\rho\rangle=1/2$, $J_{\rm num}$
is current through toroid of width $N_w$, length $N_r$ rings ($N_s=N_w \times 4N_r$), 
effective bond rates $p=1$ on bonds parallel to flow direction,, $p=1/2$ otherwise, 
as given  by numerical simulations with $N_{\rm sam}=10^{\,6}$, $N_{\rm set}=10$ 
(see text);
$J_{\rm 1d}$ is current through one-dimensional system, given by
 Eq.~(\protect{\ref{eq:j1dpbc}})~.
}
\vskip 0.2cm
\begin{ruledtabular}
\begin{tabular}{@{}ccccc}
$\ \ N_s$ &$N_w$ &$N_r$& $J_{\rm num}$  & $J_{\rm 1d}$    \\
\hline\noalign{\smallskip}
$\ \ 80$ & $4$  & $5$  & $0.253172(74)$   &  $0.2531646\dots$ \\
$\ \ 96$ & $2$  & $12$ & $0.252635(51)$   &  $0.2526316\dots$ \\
$\ \ 96$ & $3$  &  $8$ & $0.252641(57)$   &  $0.2526316\dots$ \\
$\ \ 96$ & $4$  &  $6$ & $0.252656(45)$   &  $0.2526316\dots$ \\
$\ \ 96$ & $6$  &  $4$ & $0.252626(49)$   &  $0.2526316\dots$ \\
$\ \ 96$ & $8$  &  $3$ & $0.252645(61)$   &  $0.2526316\dots$ \\
$\ \ 96$ &$12$  &  $2$ & $0.252636(46)$   &  $0.2526316\dots$ \\
$\ \ 160$& $4$ &  $10$ & $0.251589(39)$   &  $0.2515723\dots$ \\
\end{tabular}
\end{ruledtabular}
\end{table}

We checked the predictions associated with Eq.~(\ref{eq:rates1}) for a 
nanotube with open boundary conditions at the ends, at selected locations on the
$\alpha$--$\beta$ phase diagram; see Table~\ref{t4}. Agreement with mean-field 
theory is very good at $(\alpha,\beta)=(1/2,1/2)$; at $(1/4,1/4)$ and, 
especially, at $(\alpha,\beta)=(1,1)$, differences between finite-lattice 
numerical results for the nanotube and the corresponding exact ones for the 
chain are somewhat significant (up to $5\%$ for $N_s=104$ at the latter point) 
for small systems; however, they tend to vanish as $N_s$ increases. 
\begin{table}
\caption{\label{t4}
For systems with  open boundary conditions, $N_s$ sites, and $(\alpha,\beta)$ as 
specified, $J_{\rm num}$
is current through nanotube of width $N_w$, length $N_r$ rings ($N_s=N_w \times 
(4N_r+1)$), 
effective bond rates $p=1$ on bonds parallel to flow direction,, $p=1/2$ otherwise, 
as given  by numerical simulations with $N_{\rm sam}=10^{\,6}$, $N_{\rm set}=10$ 
(see text);
$J_{\rm 1d}$ is current through one-dimensional system, see e.g.
Refs.~\protect{\onlinecite{derr93,rbsdq12}}~.
}
\vskip 0.2cm
\begin{ruledtabular}
\begin{tabular}{@{}ccccc}
$\ \ N_s$ &$N_w$ &$N_r$& $J_{\rm num}$  & $J_{\rm 1d}$    \\
\hline\noalign{\smallskip}
\multicolumn{5}{c}{$(\alpha,\beta)=(1/2,1/2)$}\\
\hline\noalign{\smallskip}
$\ \ 104$ & $8$  & $3$  & $0.249994(65)$   &  $1/4$ \\
$\ \ 200$ & $8$  & $6$ &  $0.250006(44)$   &  $1/4$ \\
$\ \ 246$ & $6$  & $10$ &  $0.250011(36)$   &  $1/4$ \\
$\ \ 390$ & $6$  & $16$ &  $0.249992(31)$   &  $1/4$ \\
\hline\noalign{\smallskip}
\multicolumn{5}{c}{$(\alpha,\beta)=(1/4,1/4)$}\\
\hline\noalign{\smallskip}
$\ \ 104$ & $8$  &  $3$ & $\!\! 0.18390(7)$   &  $0.18575\dots$ \\
$\ \ 200$ & $8$  &  $6$ & $0.18620(11)$   &  $0.186574\dots$ \\
$\ \ 246$ & $6$  &  $10$ & $0.18661(11)$  &  $ 0.186745\dots$ \\
$\ \ 390$ & $6$  & $16$ &  $\!\!0.18694(9)$  &  $0.187022\dots$ \\
\hline\noalign{\smallskip}
\multicolumn{5}{c}{$(\alpha,\beta)=(1,1)$}\\
\hline\noalign{\smallskip}
$\ \ 104$ & $8$  &  $3$ & $0.267521(56)$   &  $ 0.2535888\dots$ \\
$\ \ 200$& $8$ &  $6$ & $0.256993(40)$   &  $0.2518703\dots$ \\
$\ \ 246$& $6$ &  $10$ & $0.253495(34)$   &  $0.2515213\dots$ \\
$\ \ 390$& $6$  & $16$ &  $0.251827(34)$   &  $0.2509603\dots$ \\
\end{tabular}
\end{ruledtabular}
\end{table}

We also probed the case with all effective bond rates $p=q=1$. 
Finite-size effects were generally dealt with by considering systems
of varying widths ($N_w \lesssim 15$ rings), and lengths ($N_r \lesssim 50$
rings). Within these ranges we found that results became essentially independent 
of $N_w$; for fixed (large) $N_w$ we extrapolated the corresponding sequences of 
finite-length currents against $N_r^{-1}$ via linear, or at most quadratic, fits.  
With PBC,
we confirmed that the maximal current is found at $\langle\rho\rangle=1/2$,
consistent with theory (see Sec.~\ref{theo:pbc}). However, numerics
gives $J_c=0.3492(1)$, slightly above the prediction of Eq.~(\ref{eq:jcntpq1}).
With open boundary conditions at the ends, evaluating currents at and near
$(\alpha,\beta)=(1,1)$ again gives $J=0.3492(1)$, agreeing with the PBC
result to four significant digits.

\subsection{Ribbons} 
\label{sec:ribbon}

\begin{figure}
{\centering \resizebox*{3.3in}{!}{\includegraphics*{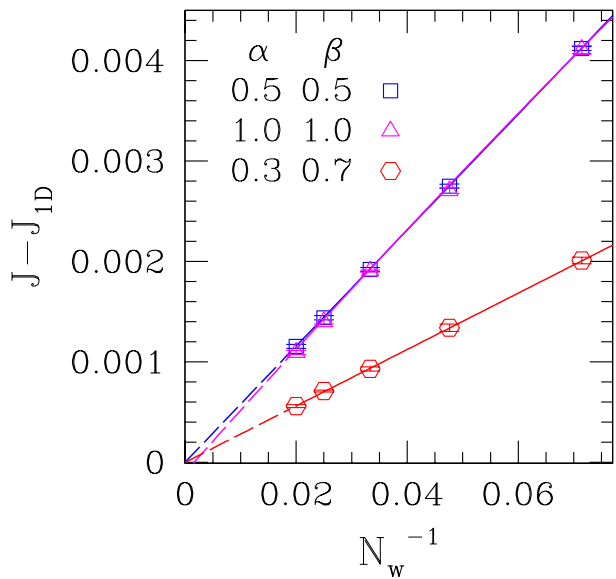}}}
\caption{(Color online) 
Simulation results for current against inverse system width $N_w^{-1}$ for nanoribbons with 
open boundary conditions, all nominal bond rates equal to unity, site update.  
Each point represents previous extrapolation to $N_r \to \infty$ at 
fixed $N_w$. $J_{1D}$ is the $N_r \to \infty$ current for one-dimensional systems
with the same injection and ejection rates: $J_{1D}=0.25$ for $(\alpha,\beta)=(0.5,0.5)$
and $(1.0,1.0)$, and $0.21$ for $(\alpha,\beta)=(0.3,0.7)$. The lines are quadratic
fits to data.
}
\label{fig:jrsu}
\end{figure}

Here, we only consider open boundary conditions at the ribbon's ends,
and all nominal bond rates are taken as unitary. Initially we use site update 
procedures, which  halves the effective rates for nearly all bonds not 
parallel to the mean flow  direction. The difference to the nanotubes of 
Sec.~\ref{sec:nanotube} [$\,$where all such bonds have their rates halved,
thus the system's rates are given by Eq.~(\ref{eq:rates1})$\,$] 
is a boundary effect: it arises because those bonds on the ribbons' edges, 
along which the flow goes 
inward, do not immediately follow a "forking" site, see Fig.~\ref{fig:nnt}.
The effective rates of such bonds then remain equal to their nominal value.
Therefore, one expects the discrepancies between steady-state currents
on ribbons and on nanotubes to vanish
as the number $N_w$ of elementary units across both systems increases.
It was predicted in Sec.~\ref{subsec:theo-mf}, and numerically
verified in Sec.~\ref{sec:nanotube}, that nanotubes with
effective bond rates given by Eq.~(\ref{eq:rates1}) behave effectively
as one-dimensional systems, so this must also be the asymptotic behavior
of ribbons. We have checked, for selected points on the
$(\alpha, \beta)$ phase diagram, that this indeed happens, only with
finite-width (and -length) effects generally more significant than for
nanotubes; see Fig.~\ref{fig:jrsu}.

We also made all effective rates equal to unity, by using bond update procedures.
We concentrated on evaluating the maximal (or critical) current across the
system, by making $\alpha=\beta=1$. For fixed width $N_w$, we produced
sequences of steady-state current estimates with growing length $N_r \leq 50$.
By fitting such sequences to parabolic forms in $N_r^{-1}$, we found that
the extrapolated values [$\,$for $N_r \to \infty\,$] $J_\infty(N_w)$  still depended
significantly on $N_w$. Finally, we extrapolated the sequence of $J_\infty(N_w)$
against $N_w^{-1}$, finding $J_c=\lim_{N_w\to \infty}J_\infty(N_w)=0.3493(1)$.

\section{Discussion and Conclusions} 
\label{sec:conc}

We have presented a mean-field theory for driven flow with exclusion
in graphene-like structures, and numerically checked its predictions
for steady state current on the necklace and nanotube structures,
with both PBC and open boundary conditions at the ends. 

For all bond rate combinations in which the mean field mapping reduces 
to the chain case,  currents on necklaces and nanotubes match those in 
the strictly one-dimensional systems. For PBC this includes finite-size 
effects, see Tables~\ref{t1} and~\ref{t3}. For open boundary conditions,
the absence of size dependence on the factorizable line $\alpha+\beta=1$ 
is reproduced; away from that line, finite-system corrections slightly differ 
from 1D ones, but discrepancies die away as system size increases
(see Tables~\ref{t2} and~\ref{t4}).

With bond rates such that no reduction to the chain case occurs, 
the maximal (critical) currents $J_c^{\,\rm MF}$ predicted by mean field 
theory [$\,$see Eqs.~(\ref{eq:necksp1}),
~(\ref{eq:necksp2}),~(\ref{eq:jcntpq1})$\,$], 
appear to be slightly off numerical results, $J_c^{\,\rm num}$ 
(at most by $2-3\%$).
Interestingly, for the cases just mentioned, one has  
$J_c^{\,\rm MF} > J_c^{\,\rm num}$ for the necklace, while
$J_c^{\,\rm MF} < J_c^{\,\rm num}$ for the nanotube.

Symmetry, or lack thereof, of the fundamental current-density relationship 
for PBC and general bond rates is correctly predicted by mean field theory 
(necklace and nanotube), see Sec.~\ref{theo:pbc}.

In this and the following paragraph, we only refer to
the special case with all nominal bond rates equal to unity. We found in 
Sec.~\ref{subsec:neck-num}
that the maximal (critical) current on the necklace structure is 
$J_c=0.3233(3)$ for site update, and $J_c=0.395(1)$ for bond update, 
to quote an aggregate of the results given there. For nanotubes, 
the mean-field theory of Sec.~\ref{subsec:theo-mf} predicts
equivalence to the one-dimensional TASEP for the bond rates
quoted in Eq.~(\ref{eq:rates1}). Such rates are effectively reproduced
by using site update with all nominal bond rates unitary. So, if the
actual transport mechanism on zigzag CNTs displays the characteristics of 
site update, one would expect such structures to behave as effectively
(rather than quasi-) one-dimensional. 

\begin{figure}
{\centering \resizebox*{3.3in}{!}{\includegraphics*{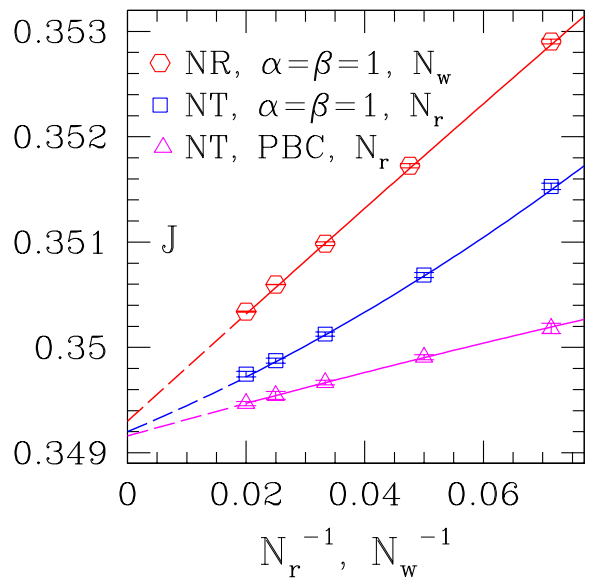}}}
\caption{(Color online) 
For all effective bond rates equal to unity,
current against inverse system size $N_r^{-1}$ (triangles, squares) or $N_w^{-1}$
(hexagons; each point represents previous extrapolation to $N_r \to \infty$ at 
fixed $N_w$). Hexagons: nanoribbons (NR) with open boundary conditions, 
$\alpha=\beta=1$. Squares: nanotubes (NT) with open boundary conditions,
$\alpha=\beta=1$. Triangles: NT with PBC, $\langle \rho \rangle =1/2$
(corresponding to maximal current, see Sec.~\ref{theo:pbc}).  
Points correspond to simulations. }
\label{fig:jall_r1}
\end{figure}

For nanoribbons, no mean-field theory has been developed here,
for reasons explained at the end of Sec.~\ref{subsec:ext}. However,
the considerations of Sec.~\ref{sec:ribbon} show that broad ribbons
should carry the same current as nanotubes  
(although finite-size effects can be rather significant), so e.g. for
site update  and
$N_w \gg 1$ the maximal current on both structures approaches $J_c=1/4$.
This is verified numerically, as illustrated in Fig.~\ref{fig:jrsu}.
For bond update, our data are summarized in Fig.~\ref{fig:jall_r1}, which
pertains both to Sec.~\ref{sec:nanotube} and to Sec.~\ref{sec:ribbon}. 
Fig.~\ref{fig:jall_r1} strongly suggests that both types of structure 
asymptotically support the same maximal current also when all effective
rates are equal.
We quote $J_c=0.3492(2)$, allowing for the uncertainties of all three sequences of
estimates displayed there.

Regarding the applicability of the generalized TASEP model discussed
here to experimentally realized systems, we first note that transport in  
CNT and CNR is predominantly governed by band electrons~\cite{RMP}, 
for which neither a bond nor a site can be uniquely assigned at any time-step.
Nevertheless, residual signatures of the topology of the hexagonal skeleton
can be expected to remain, such as the trend followed by (normalized)
current against increasing width, seen in Figs.~\ref{fig:jrsu} and~\ref{fig:jall_r1}.
On the other hand, finer details of the TASEP behavior unveiled here possibly have no 
discernible counterparts in such C-based materials.

Turning now to quantum dot (QD) systems, site update would be adequate, e.g., to 
model QD arrays in which 
the electron remains bound to a specific QD for dwell times much longer than the
inverse hopping attempt rate onto a neighboring QD. 

One might consider, e.g., a honeycomb arrangement of 
QDs, so that each QD may be empty or occupied by one electron, while
a second electron is excluded by Coulomb blockade. Electrostatically defined
QD arrays have already been fabricated via electrodes over a
two-dimensional electron gas on a semiconductor/barrier 
interface~\cite{IvW11}. 
Recent experimental investigations of electron hopping transport
in systems of self-assembled QD chains indicate that such studies in more complex
geometries may soon be accessible~\cite{jap13}.

Of course planar arrangements of QDs provide a
physical realization of a nanoribbon, but not a nanotube. 
For a QD array forming a
ribbon-shaped cutout (along a given bond direction) of the
honeycomb lattice, individual tunnel barrier strengths could be tuned by the
electrodes shaping the confining potential, thus defining the
bond rates. The various bond rates considered here would
then be experimentally accessible.
Bond update procedures would be applicable to strongly covalent arrays, and may be
useful to the study of transport in macromolecules.

Finally we mention preliminary investigations for open boundary conditions at the ends, 
regarding  the predictions given in 
Eqs.~(\ref{eq:acbc1})--(\ref{eq:acbc3}),~(\ref{eq:bdy2b}),~(\ref{eq:bdy3}),
~(\ref{eq:line}), on the location and properties of the critical points, 
phase boundaries, and coexistence and factorization lines. We focus on the
issue of factorization. Among numerical tests of various degrees of factorization 
we have looked at:\par\noindent
\ {\it (i)} constancy of sublattice density profiles,\par\noindent
\ {\it (ii)} satisfaction of Eq~(\ref{eq:bdy3}), and\par\noindent
\ {\it (iii)} factorization of correlation functions
in the steady state, both on, and off, predicted factorization lines. 

For the latter we evaluated 
\begin{equation}
C_{ij} \equiv \langle J_{ij}\rangle - p_{ij}\, \langle \tau_i \rangle\,\left( 1 - 
\langle\tau_j\rangle\right)\ ,
\label{eq:c_ij}
\end{equation}
where the average current $\langle J_{ij}\rangle$ across a chosen bond $ij$ 
with rate $p_{ij}$ is the correlation function which, if factorizing, 
makes the quantity $C_{ij}$ vanish.

The severe test $(iii)$ has been very informative. For example, for the
nanotube it shows vanishing of $C_{ij}$  to the accuracy of simulation
(typically $1$ part  in $10^5$) in the case with $p=1$, $q=1/2$ on (and only on) the
predicted line Eq.~(\ref{eq:line}); this is a non-trivial higher-dimensional
generalization of a well known result for the linear chain. On the other
hand, for other cases such as  $p=1=q=1$, in simulations of similar
accuracy, the factorization is no better than $1$ part in $10^2$. These results 
apply whether or not extra stochastic fluctuations  (such as those which
distinguish the steady state current from the current activity~\cite{rbsdq12})
are included.

The open and relevant issue of factorization in these systems deserves full
attention in its own right. A complete discussion, complementing the
present study and including comprehensive numerical results and theory
based on the hierarchy of equations of motion and on matrix representations of 
the Master Equation, will be presented elsewhere.

\begin{acknowledgments}
We thank F. H. L. Essler, F. Pinheiro, and R. B. Capaz for interesting discussions.
S.L.A.d.Q. thanks the Rudolf Peierls Centre for Theoretical Physics, Oxford, for
hospitality during his visit. The research of S.L.A.d.Q., M.A.G.C, and B.K. is supported by 
the Brazilian agencies CNPq  (Grants Nos. 302924/2009-4, 302040/2009-9, and 160714/2011-7), 
and FAPERJ (Grants Nos. E-26/101.572/2010, E-26/102.760/2012, and E-26/110.734/2012).
\end{acknowledgments}

\end{document}